\documentclass[twocolumn]{aastex6}
\usepackage{graphicx}
\usepackage[countmax]{subfloat}
\usepackage{booktabs}
\usepackage{mwe}
\usepackage{amsmath}
\newcommand{\RNum}[1]{\uppercase\expandafter{\romannumeral #1\relax}}

\begin{document}
\title{Optical Flux and Spectral Variability of the TeV Blazar PG 1553$+$113}
\author{Ashwani Pandey\altaffilmark{1,2}, Alok C. Gupta\altaffilmark{1}, Paul J. Wiita\altaffilmark{3}, \& S. N. Tiwari\altaffilmark{2}}

\altaffiltext{1}{Aryabhatta Research Institute of Observational Sciences (ARIES), Manora Peak, Nainital 263002, India; ashwanitapan@gmail.com}
\altaffiltext{2}{Department of Physics, DDU Gorakhpur University, Gorakhpur 273009, India; acgupta30@gmail.com}
\altaffiltext{3}{Department of Physics, The College of New Jersey, 2000 Pennington Rd., Ewing, NJ 08628-0718, USA; wiitap@tcnj.edu}


\begin{abstract}
We present the results of our optical (VRI) observations of the TeV blazar PG 1553$+$113 over eight nights in 2016 April. We monitored the blazar quasi-simultaneously in V and R bands each night and examined the light curves (LCs) for intraday flux and color variations using two of the most powerful tests; the power-enhanced {\it F}-test and the nested ANOVA test. The source was found to be significantly ($>99\%$) variable in both V and R bands only on April 13, while clear variations only in R band LCs were seen on April 8 and 12. No temporal variation was seen in the color during the observation period. We did not find any significant correlation between V-R color index and R magnitude on any observing night.
We found a mean optical spectral index of $\sim 0.83 \pm 0.02 $ with a maximum variation of 0.21 by fitting a power law ($F_{\nu} \propto \nu^{-\alpha}$) in the optical (VRI) spectral energy distribution of  PG 1553$+$113. We briefly discuss the possible physical processes responsible for the observed flux and spectral variability. 
\end{abstract}


\section{INTRODUCTION} \label{sec:intro}
In the orientation based unification scheme of active galactic nuclei (AGNs), blazars are the AGNs having relativistic jets aligned at an angle of  $ \leq 10^{\circ}$ from the observer's line of sight \citep{1995PASP..107..803U}. The two sub-classes of blazars are BL Lacertae objects (BL Lacs), which are characterized by essentially featureless optical spectra, and flat spectrum radio quasars (FSRQs), which have broad emission lines in their optical spectra \citep[e.g.][]{1991ApJS...76..813S, 1996MNRAS.281..425M}. Their non-thermal radio to $\gamma-$ray spectral energy distributions (SED) exhibit  characteristic double bump structures in a $\nu F_{\nu}-\nu$ representation \citep{1998MNRAS.299..433F}. The lower energy component, which peaks at IR to optical frequencies in the low-frequency peaked blazars (LBLs) and at FUV to X-ray frequencies in the high-frequency peaked blazars (HBLs), is attributed to synchrotron emission of relativistic electrons in the blazar jet. The high energy bump, peaking at GeV energies in LBLs and at TeV energies in HBLs, is produced  either due to inverse Compton scattering of the low energy photons by the same electrons responsible for the synchrotron emission (leptonic model; e.g., \cite{2007Ap&SS.307...69B}) or via emission  arising from relativistic protons (hadronic model; e.g., \cite{2003APh....18..593M}).

PG 1553$+$113 (1ES 1553$+$113; $\alpha_{\rm 2000}$ = 15h55m43.0s; $\delta_{\rm 2000} = +11^{\circ}11^{\prime}24.4^{\prime\prime}$) 
was first identified as a BL Lac object in the Palomar--Green survey of ultraviolet-excess stellar objects \citep{1986ApJS...61..305G}. It is classified as a BL Lac object due to its featureless optical spectrum \citep{1983BAAS...15..957M} and significant ($\Delta m \sim 1.9$ mag) optical variability \citep{1988ESASP.281b.303M}.
It is a high frequency peaked BL Lac object (HBL) \citep{1990PASP..102.1120F} at a red-shift of z $\sim 0.5$ \citep{2010ApJ...720..976D, 2015ApJ...802...65A}. 
The ratio of X-ray to radio flux (log($F_{2keV}/F_{5GHz}$)) of PG 1553$+$113 ranges from $-$4.37 to $-$3.88 \citep{2006AJ....132..873O}, suggesting that it is an extreme HBL \citep{2003AJ....125.1060R}. Its optical spectral index was found to be nearly constant ($\alpha \sim -1\footnote{using the convention, $F_{\nu} \propto \nu^{\alpha}$}$) with a maximum variation of 0.24 during 1986--1991 by \cite{1994ApJS...93..125F}. It was observed in the bright state with average R band magnitude of $\sim$13.3 during March--August 2010 by \cite{2012MNRAS.425.3002G}. In a recent multi-band optical study of TeV blazars, PG 1553$+$113 was detected with $m_R \sim 13.81-14.40$ and $m_V \sim 14.17-14.71$ \citep{2016MNRAS.458.1127G}.

To date, about 65 TeV blazars\footnote{\url{http://tevcat.uchicago.edu}} have been detected, most ($\sim48$) of which are HBLs.
PG 1553$+$113 was discovered as a TeV HBL with $\gamma$-ray photon index of $\Gamma = 4.0 \pm 0.6$ by H.E.S.S. \citep{2006A&A...448L..19A} and has been studied from radio to $\gamma$-rays in different observation campaigns \citep[e.g.,][]{2006AJ....132..873O,2015MNRAS.454..353R,2015ApJ...813L..41A,2017MNRAS.466.3762R}. 
It has gained additional attention after the claim of a $2.18 \pm 0.08$ years quasi-periodicity in its $\gamma-$ray flux \citep{2015ApJ...813L..41A}. To explain such periodicities, of the timescale of few years, a model based on  periodic procession of jet in a binary system of super-massive black holes was proposed  \citep{2017MNRAS.465..161S,2017ApJ...851L..39C}.

Flux variability at diverse timescales is one of the characteristic properties of blazars. On the basis of timescales of occurrence, variability can be divided into three classes; intraday variability (IDV) or micro-variability (occurring on a timescale of minutes to hours), short term variability (STV)  (taking place on a timescale of days to months), and long term variability (LTV) (over a timescale of several months to years) \citep[e.g.,][]{1995ARA&A..33..163W,2004A&A...422..505G}.
The first clear optical IDV detection was reported by \cite{1989Natur.337..627M} in the light curves of BL Lac objects. Since then the optical variability of blazars on diverse timescales has been studied extensively \citep[e.g.,][]{1990PhDT........11C, 1996A&A...305...42H, 1998A&AS..132...83B, 2001A&A...369..758F, 2002MNRAS.329..689X, 2005MNRAS.356..607S, 2008AJ....135.1384G, 2008NewA...13..375G, 2012MNRAS.420.3147G, 2012MNRAS.424.2625B, 2012MNRAS.425.3002G, 2015MNRAS.450..541A, 2015MNRAS.452.4263G, 2016MNRAS.455..680A, 2016MNRAS.458.1127G}. In the optical regime, HBLs are found to be less variable than LBLs and their variability amplitudes are also much smaller than that of LBLs \citep{1994ApJ...428..130J}.
Flux variations at optical frequencies are often accompanied by color variations. It has been found that the BL Lac objects, in general, follow a bluer-when-brighter (BWB) trend while the FSRQs tend to follow a redder-when-brighter (RWB) trend \citep[e.g.,][]{2000ApJ...537..101F, 2012AJ....143..108W, 2015A&A...573A..69W}. 

As the intraday variability is thought to originate from the compact emission regions that are close to the central super-massive black hole, the  study of IDV provides an opportunity to understand the physics and geometry of these otherwise inaccessible inner regions. The main motivation of this work is to study the optical flux and spectral variations of the TeV blazar PG 1553$+$113 on IDV timescales. Here we report the optical photometric observations of the HBL PG 1553$+$113 on IDV and STV timescales using two Indian telescopes during 2016 April 06--16. We also investigated the 
color variations and the spectral variations using optical SEDs during our monitoring.

The paper is structured as follows: Section \ref{sec:data} describes the details of observations and data reduction; in Section \ref{sec:analysis} we discuss the analysis techniques used. Results of our flux and spectral variability studies are given in Section \ref{sec:res} and Section \ref{sec:diss} presents a discussion of our results and our conclusions.

\section{OBSERVATIONS AND DATA REDUCTION} \label{sec:data}
We have observed the TeV blazar PG 1553$+$113 on 2016 April 6 and 16 with the 1.30 m (f/4) Devasthal Fast Optical Telescope (DFOT) in {\it B, V, R,} and {\it I} filters, and from 2016 April 8 through April 13 with the 1.04 m (f/13) Sampuranand Telescope (ST) in {\it V, R,} and {\it I} filters at the Aryabhatta Research Institute of Observational Sciences (ARIES), Nainital, India. Both of these telescopes are Ritchey-Chretien (RC) reflectors with  Cassegrain focus and we used Johnson {\it UBV} and Cousins {\it RI} filters. The technical details of these two telescopes and the instruments used for observations are given in Table \ref{tab:1}. The source was observed for a total of 8 nights with quasi-simultaneous observations in {\it V} and {\it R} bands every night. The bias frames were taken regularly throughout the observation and the sky flats in each filter were obtained during twilight. The observation log of optical photometric observations of PG 1553$+$113 is given in Table \ref{tab:obs_log}.

The pre-processing of the raw data, which involves bias subtraction, flat fielding and cosmic ray removal, was performed using the standard routines of the Image Reduction and Analysis Facility (IRAF\footnote{IRAF is distributed by the National Optical Astronomy Observatory, which is operated by the Association of Universities for Research in Astronomy (AURA) under a cooperative agreement with the National Science Foundation.}). The data then was processed using the Dominion Astronomical Observatory Photometry (DAOPHOT \RNum{2}) software to obtain the instrumental magnitudes of the blazar PG 1553$+$113 and the stars in the image frames by applying the aperture photometry technique, using the APPHOT routine. Aperture photometry was performed in each image frame using four different 
concentric aperture radii, i.e., $\sim$ 1 $\times$ Full Width at Half Maximum (FWHM), 2 $\times$ FWHM, 3 $\times$ FWHM, and 4 $\times$ FWHM. However, it was found that the aperture radius 2 $\times$ FWHM always provided the best signal-to-noise (S/N) ratio, so we used that aperture radius for our final results \citep[e.g.,][]{2015A&A...582A.103G}. During each observation, three or more local standard stars were observed in the same blazar field.
Out of these three, the two standard stars (stars 2 and 3 from Fig.\ 1 of \cite{2015MNRAS.454..353R}) having magnitude and  color closer to that of the blazar were used to check the mutual non-variability of those standard stars. Finally one comparison star (star 2) was used to calibrate the instrumental magnitudes of the TeV blazar PG 1553$+$113. Since the blazar and the comparison star 2 were both observed in the same frame no atmospheric extinction corrections were performed. The photometric data of our observations are provided in Table \ref{tab:data}.

\begin{table}[!h]
\caption{Details of telescopes and instruments used}             
\label{tab:1}                   
\centering   			
\resizebox{0.5\textwidth} {!}{                    
\begin{tabular}{l c c}          
\hline                		
            & A                        & B \\   		
\hline                           
Telescope   		    & 1.30 m DFOT  	& 1.04 m ST  \\           
CCD Model   		    & Andor 512 CCD	       			& Tektronics 1K CCD     \\
Chip Size (pixels) 	    & $512 \times 512$          		& $1024 \times 1024$        \\
Pixel Size ($\mu m$)        &  $16 \times 16$				& $24 \times 24$   \\
Scale (arcsec/pixel)   	    &  0.64					& 0.37    \\ 
Field ($arcmin^2$)    	    & $5.5 \times 5.5$				& $ 6 \times 6 $      \\ 
Gain ($e^-$/ADU)	    &  1.4					& 11.98     \\ 
Read-out Noise ($e^-$ rms)  &  6.1					& 6.9     \\ 
Typical Seeing (arcsec)     &  1.3--3.2					& 1.4--2.6     \\ 
\hline                           
\end{tabular}
}\\
\end{table}

\begin{figure*}
\centering
\includegraphics[width=19cm, height=10cm]{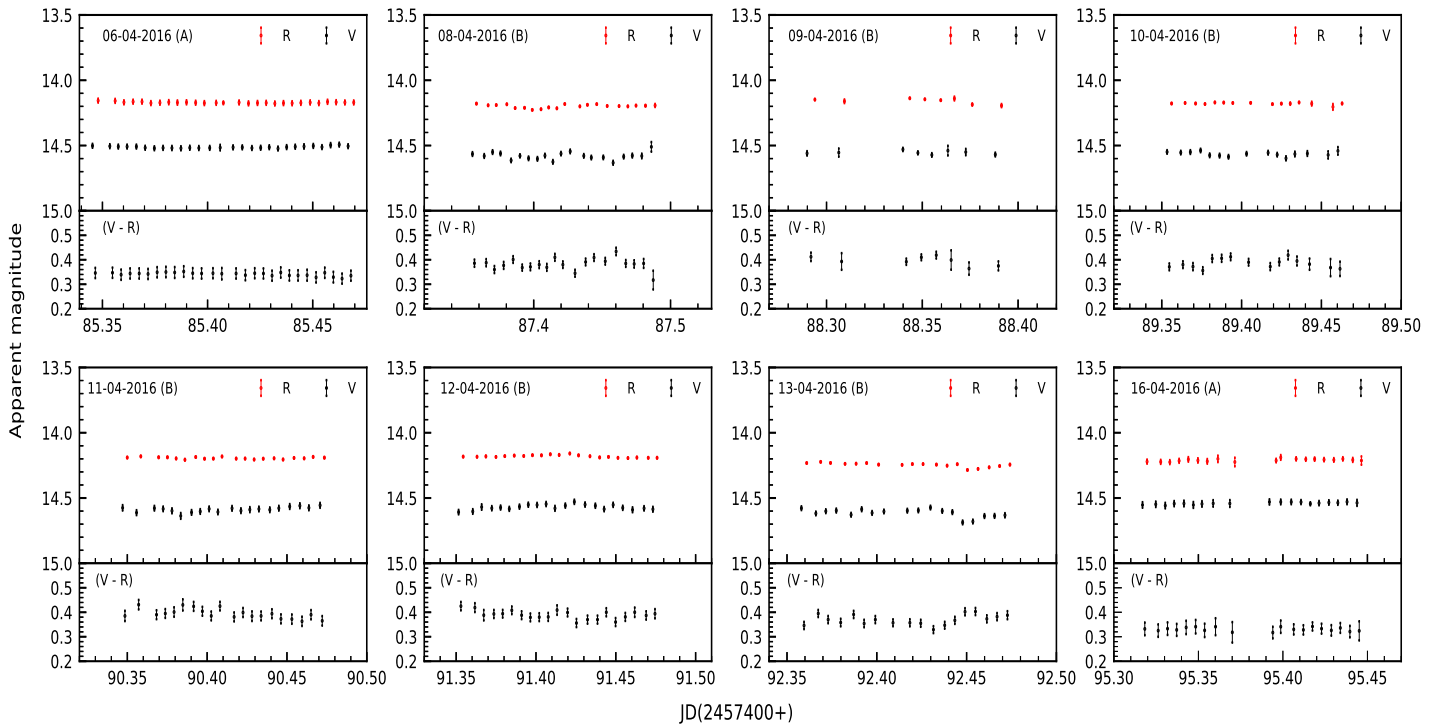}
\caption{Upper panel in each plot shows  the optical light curves of TeV blazar PG 1553$+$113; red denotes the R filter and black denotes V filter. In the bottom panel the color (V$-$R) variation on IDV timescales is shown. The observation dates and the telescope used (in parentheses) are displayed.} 
\label{fig:lc}

\end{figure*}

\begin{table}[!h]
\caption{Observation log for PG 1553$+$113}            
\label{tab:obs_log}                   
\centering     
                 
\begin{tabular}{l c c}           
\hline                		 
Observation date & Telescope  & Data points \\
yyyy-mm-dd   &            &  B, ~V, ~R, I           \\		 
\hline                          
2016-04-06	&  A  &   1, 28, 28, 1\\
2016-04-08	&  B  &   0, 20, 20, 1\\
2016-04-09	&  B  &   0, ~8, ~8, 1\\
2016-04-10	&  B  &   0, 16, 16, 1\\
2016-04-11	&  B  &   0, 20, 20, 1\\
2016-04-12	&  B  &   0, 21, 21, 1\\
2016-04-13	&  B  &   0, 18, 18, 1\\
2016-04-16	&  A  &   1, 19, 19, 1\\ 
\hline                          
\end{tabular}
\end{table}

\begin{table}
\caption{Photometric data (R band) of the TeV blazar PG 1553$+$113.}            
\label{tab:data}                   
\centering     
\hskip-1.cm                
\begin{tabular}{ccc}           
\hline                		 
JD & Magnitude &  Error  \\
\hline                          
2457485.347963 & 14.156 &  0.015 \\
2457485.356053 & 14.158 &  0.015 \\
2457485.360220 & 14.169 &  0.015 \\
2457485.364583 & 14.164 &  0.015 \\
2457485.368819 & 14.164 &  0.015 \\
2457485.373044 & 14.175 &  0.015 \\
2457485.377338 & 14.172 &  0.015 \\
\hline                           
\end{tabular}\\
(This is a sample data set in the R band. The complete photometric data for V and R band LCs of PG 1553$+$113 are available in the online journal.)
\end{table}

\section{ANALYSIS TECHNIQUES} \label{sec:analysis}
We have examined the differential light curves of the blazar PG 1553$+$113 for intraday variations using the power-enhanced {\it F}-test and the two-stage nested analysis of variance (ANOVA) test, or simply nested ANOVA test, which have been argued to be more reliable and powerful than the widely used statistical tests such as  {\it C}-test and the standard {\it F}-test \citep{2014AJ....148...93D, 2015AJ....150...44D}. The key idea of both the probes is to include the light curves of several comparison stars in the analysis which increases the power of the probe. 

\subsection{Power-enhanced {\it F}-test}\label{sec:f_test}
In the power-enhanced  {\it F}-test, we compare the blazar light curve variance to the combined variance of multiple comparison stars. This test has been used in several recent studies for detecting microvariations in blazar light curves \citep[e.g.,][]{2015MNRAS.452.4263G, 2016MNRAS.460.3950P, 2017MNRAS.466.2679K}.
The power-enhanced {\it F}-statistic is defined as \citep{2014AJ....148...93D}: 
\begin{equation} 
F_{enh} = \frac{s_{blz}^2}{s_c^2},
\end{equation} 
where
\begin{equation}
s_c^2 = \frac{1}{(\sum_{j=1}^k N_j)-k}\sum_{j=1}^{k} \sum_{i=1}^{N_i} s_{j,i}^2,
\end{equation}
and $s_{blz}^2$ is the variance of the (blazar -- reference star) differential instrumental light curves (DLCs) and $s_c^2$ is the combined variance of the (comparison star -- reference star) DLCs; $N_j$ is the number of observations of the $j$th comparison star and $k$ is the total number of comparison stars. The scaled square deviation, $s_{j,i}^2$, for the $j$th comparison star is calculated as
\begin{equation}
s_{j,i}^2 = \omega_j(m_{j,i}-\bar{m_j})^2,
\end{equation}
where $\omega_j$ is a scaling factor used to scale the variance of $j$th comparison star to the level of the blazar \citep{2011MNRAS.412.2717J}, $m_{j,i}$ is the differential magnitude and  $\bar{m_j}$ is the mean magnitude of the comparison star DLC.

In our case, we have three comparison field stars (S2, S3 and S4) from which S2, having magnitude closest to the blazar, is taken as the reference star. Hence we have two ($k = 2$) field stars as comparison stars. Since all the comparison stars and the blazar have the same number of observations ($N$), the number of degrees of freedom in the numerator and denominator in the {\it F}-statistics are $\nu_1 = N-1$ and $\nu_2 = k(N-1)$ respectively.
We then estimated the $F_{enh}$ value using Equation (1) and compared it with the critical value ($F_c$) at $\alpha = 0.01$ which corresponds to a confidence level of 99 per cent. A light curve is considered as variable (V) if $F_{enh} \geq F_c$, otherwise we call it non-variable (NV). 

\subsection{Nested {\it ANOVA}}\label{sec:anova}
The ANOVA test compares the means of dispersion between the groups of observations. The nested ANOVA test is an updated ANOVA test which uses several stars as reference stars to generate different differential light curves of the blazar. In contrast to power-enhanced {\it F}-test, no comparison star is needed in the nested ANOVA test, so the number of stars in the analysis has increased by one. 

In our case, we have used three reference stars (S2, S3 and S4) to generate differential LCs of the blazar. These three differential LCs are, then, divided into a number of groups with four points in each group. Following equation (4) of \cite{2015AJ....150...44D}, we calculated the mean square due to groups ($MS_G$) and mean square due to nested observations in groups ($MS_{O(G)}$). The ratio $F = MS_{G}/MS_{O(G)}$
 follows an {\it F} distribution with $a - 1$ and $a(b - 1)$ degrees of freedom, in the numerator and denominator, respectively.
For a significance level of $\alpha$ = 0.01, if the $F-$statistic $\geq$ the critical value $(F_c)$, the light curve is taken as variable (V), otherwise as non-variable (NV). 

The results of the $F_{enh}-$tests and nested ANOVA tests are presented in Table \ref{tab:var_res}.  In it, a light curve is declared as variable (V) only if significant variations were detected by both the tests, otherwise we conservatively label it non-variable (NV). 

\subsection{Intraday Variability Amplitude}
For the light curves that are found variable, we calculated the IDV amplitude (Amp) in per cent using the relation given by \cite{1996A&A...305...42H}.
\begin{equation}
Amp = 100\times \sqrt{(A_{max}-A_{min})^2 - 2 \sigma^2},
\end{equation}
where $A_{max}$ and $A_{min}$ are the maximum and minimum magnitudes, respectively, in the calibrated light curves of the blazar, while $\sigma$ is the mean error.

\section{RESULTS}\label{sec:res}
\subsection{Flux Variability} \label{sec:flux}

We monitored the blazar PG 1553$+$113 for a total of 8 nights from 6--16 April, 2016. During each night we observed the source quasi-simultaneously in V and R bands for a duration of $\sim 2-4$ hours to investigate the intraday variability properties. Single I band measurements were also made on each night and single B band measurements were made on the first and last nights.  The calibrated V and R band IDV light curves of the blazar PG 1553$+$113 are shown in the upper panel of each plot in Fig. \ref{fig:lc}.
Visual inspection of the light curves appears to show intraday variations on a couple of nights.

In order to statistically examine the V and R band light curves for intraday variations, we performed the power-enhanced {\it F}-test and the nested ANOVA test, discussed in section \ref{sec:f_test} and section \ref{sec:anova}, respectively. The results of the statistical analysis are given in Table \ref{tab:var_res}. Significant intraday variations were detected in both V and R band light curves of PG 1553$+$113  only on April 13, while no significant IDV was observed at any band on April 6, 9, 10, 11, and 16. We also found significant variations in R band light curves on April 8 and 12, although any variability in the V band on those nights was not significant. However, notice that the errors in V band light curves are roughly twice as large, and  therefore reduce the likelihood of detecting any small variations that might be present.  

We also estimated the intraday variability amplitudes for the confirmed variable LCs, shown in the last column of Table \ref{tab:var_res}, using Equation (8). The detected variability amplitude was smallest ($3.44 \%$) in R band on April 12, while the largest ($11.23 \%$) variation was observed in V band on April 13. Usually the blazar variability amplitude is larger at higher frequencies, as was seen in the one night for which both were detectable, which suggests that the blazar spectrum gets steeper with decreasing brightness and flatter with increasing brightness \citep[e.g.,][]{1998MNRAS.299...47M, 2015MNRAS.450..541A}. However, on some occasions the variability amplitude of blazars at lower frequencies was found comparable to or even larger than that at higher frequencies \citep[e.g.,][]{2000ApJ...537..638G, 2015MNRAS.452.4263G}. 

\begin{figure}
\centering
\includegraphics[width=8cm, height=8cm]{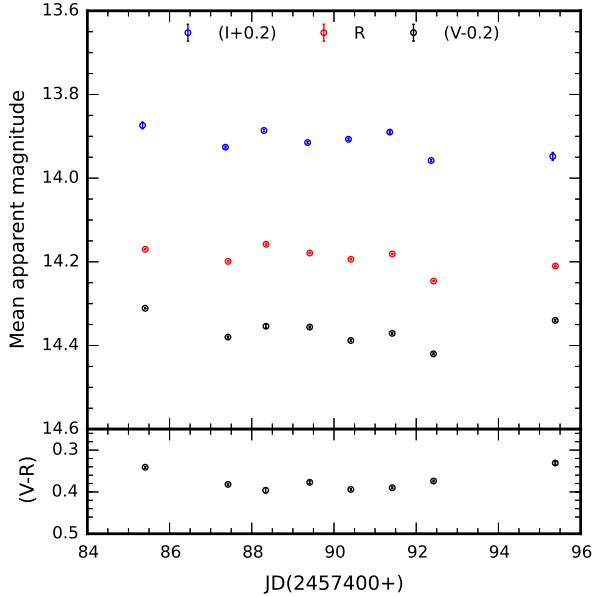}
\caption{Upper panel displays STV optical (VRI) light curves of PG 1553$+$113; they are shown in black  (V); red (R) and blue (I), respectively;  bottom panel represents the color (V$-$R) variation on STV timescales.} 
\label{fig:stv}
\end{figure}

The STV light curves of PG 1553$+$113 in V, R and I bands for the entire monitoring period are plotted in the upper panel of Fig.\ \ref{fig:stv}, where we have plotted the nightly averaged magnitudes with respect to time.
The V and I band light curves are shifted by $-0.2$ and $+$0.2 magnitudes, respectively, to make the variability pattern visible. During our monitoring period the source was detected in the brightest state of $R_{mag} = 14.138$ on April 9, while the faintest magnitude detected was $R_{mag} = 14.285$ on April 13. The mean magnitudes were 14.563, 14.193, and 13.713 in V, R, and I bands, respectively. The variability on STV timescales can be clearly seen at all three optical wavelengths.

\begin{table*}[]
\caption{Results of IDV analysis of PG 1553$+$113}            
\label{tab:var_res}                   
\centering 
\hskip-1.5cm                     
\begin{tabular}{lcccccccccc}           
\hline                		 
Observation date & Band & \multicolumn{3}{c}{{\it Power-enhanced  F-test}} & \multicolumn{3}{c}{{\it Nested ANOVA}} &  Status & Amplitude\\
\cmidrule[0.03cm](r){3-5}\cmidrule[0.03cm](r){6-8} yyyy-mm-dd & & DoF($\nu_1$,$\nu_2$ ) & $F_{enh}$ & $F_c$  & DoF($\nu_1$,$\nu_2$ ) & $F$ & $F_c$ & &$\%$ \\
\hline
20160406  & V     &  27, 54 & 1.33 & 2.11 	&   6, 21 & ~7.14  & ~3.81  & NV  &  - \\ 
	  & R     &  27, 54 & 1.34 & 2.11 	&   6, 21 & 10.27  & ~3.81  & NV &  - \\ 
	  & V-R   &  27, 54 & 1.11 & 2.11 	&   6, 21 & ~1.99  & ~3.81  & NV &  - \\ 
20160408  & V     &  19, 38 & 2.07 & 2.42 	&   4, 15 & ~1.17  & ~4.89  & NV &  - \\ 
	  & R     &  19, 38 & 4.32 & 2.42 	&   4, 15 & 15.02  & ~4.89  & V  & ~4.76\\ 
	  & V-R   &  19, 38 & 1.46 & 2.42 	&   4, 15 & ~1.60  & ~4.89  & NV &  - \\ 
20160409  & V     &  ~7, 14 & 0.24 & 4.28 	&   1, ~6 & ~3.23  & 13.75  & NV &  - \\ 
	  & R     &  ~7, 14 & 7.17 & 4.28 	&   1, ~6 & ~1.11  & 13.75  & NV &  - \\ 
	  & V-R   &  ~7, 14 & 0.44 & 4.28 	&   1, ~6 & ~7.58  & 13.75  & NV &  - \\ 
20160410  & V     &  15, 30 & 1.05 & 2.70 	&   3, 12 & ~1.34  & ~5.95  & NV &  - \\ 
	  & R     &  15, 30 & 1.09 & 2.70 	&   3, 12 & ~0.43  & ~5.95  & NV &  - \\ 
	  & V-R   &  15, 30 & 1.36 & 2.70 	&   3, 12 & ~1.42  & ~5.95  & NV &  - \\ 
20160411  & V     &  19, 38 & 0.84 & 2.42 	&   4, 15 & ~2.67  & ~4.89  & NV &  - \\ 
	  & R     &  19, 38 & 2.09 & 2.42 	&   4, 15 & ~2.40  & ~4.89  & NV &  - \\ 
	  & V-R   &  19, 38 & 1.05 & 2.42 	&   4, 15 & ~2.64  & ~4.89  & NV &  - \\ 
20160412  & V     &  20, 40 & 0.85 & 2.37 	&   4, 15 & ~7.22  & ~4.89  & NV &  - \\ 
	  & R     &  20, 40 & 4.31 & 2.37 	&   4, 15 & 14.53  & ~4.89  & V  &  ~3.44 \\ 
	  & V-R   &  20, 40 & 0.70 & 2.37 	&   4, 15 & ~6.34  & ~4.89  & NV &  -\\ 
20160413  & V     &  17, 34 & 3.57 & 2.54 	&   3, 12 & ~7.25  & ~5.95  & V  &  11.23 \\ 
	  & R     &  17, 34 & 7.45 & 2.54 	&   3, 12 & ~6.55  & ~5.95  & V  & ~6.17  \\ 
	  & V-R   &  17, 34 & 1.70 & 2.54 	&   3, 12 & ~5.32  & ~5.95  & NV &  - \\ 
20160416  & V     &  18, 36 & 0.29 & 2.48 	&   3, 12 & ~0.73  & ~5.95  & NV &  - \\ 
	  & R     &  18, 36 & 0.22 & 2.48 	&   3, 12 & ~0.58  & ~5.95  & NV &  - \\ 
	  & V-R   &  18, 36 & 0.10 & 2.48 	&   3, 12 & ~0.28  & ~5.95  & NV &  - \\ 
\hline                          
\end{tabular}
\end{table*}

\subsection{Spectral Variability} \label{sec:spec}
\begin{figure*}
\centering
\includegraphics[width=19cm, height=8cm]{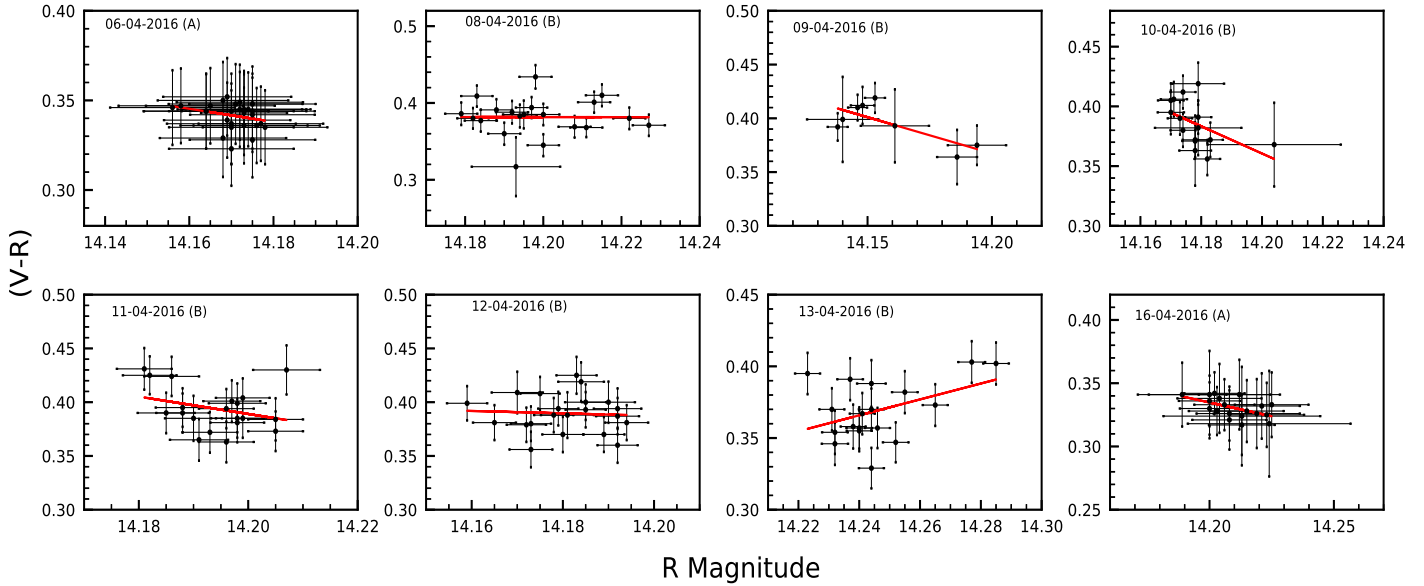}
\caption{IDV color-magnitude plots for PG 1553$+$113. The observation date and the telescope code are given in each plot. } 
\label{fig:color-mag}
\end{figure*}

Optical flux variations in blazars are often associated with spectral changes. To investigate spectral variability of the blazar PG 1553$+$113 on intranight timescales, we plotted the V$-$R color indices (CIs) with respect to time (color--time), shown in the bottom panel of Fig.\ \ref{fig:lc}, and  with respect to R band magnitude (color--magnitude), displayed for each night in Fig.\ \ref{fig:color-mag}. We have taken each pair of V, then R, band images to get the color (V$-$R) considering the time for the color measurement to be the average of the times for the V and immediately subsequent R band observations. We found no significant temporal variation in V$-$R color, as  shown in Table \ref{tab:var_res}. To investigate the color behavior of the blazar with respect to R band magnitude, we fitted a straight line of the form $CI = mR+c$ on each color--magnitude plot, the results of which are listed in Table \ref{tab:r_v}.  No significant correlation was observed between V$-$R color and R magnitude on any night, which is consistent with the non-detection of variability or detection of only low amplitude variability in the LCs. Even on April 13, when we detected variability in both V and R band LCs, the correlation coefficient is $r = 0.426$ with a corresponding substantial value of null hypothesis probability, $p = 0.078$,  indicating no clear color variation with brightness. 

\begin{figure}
\centering
\includegraphics[width=8cm, height=8cm]{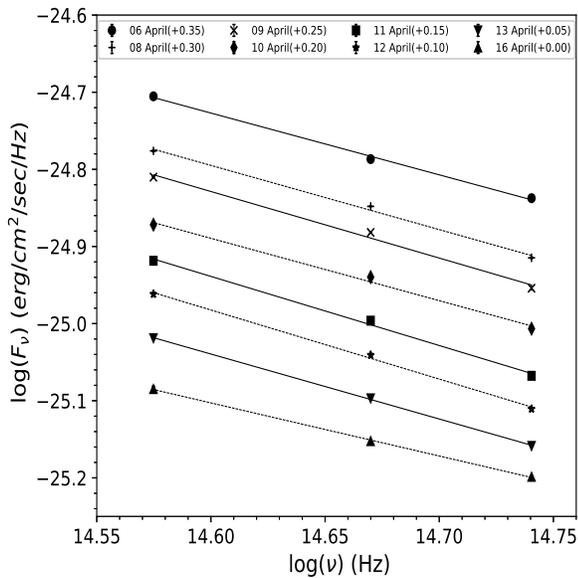}
\caption{The SED of PG 1553$+$113 in V, R, and I bands.} 
\label{fig:sed}
\end{figure}

\begin{figure}
\centering
\includegraphics[width=8cm, height=6cm]{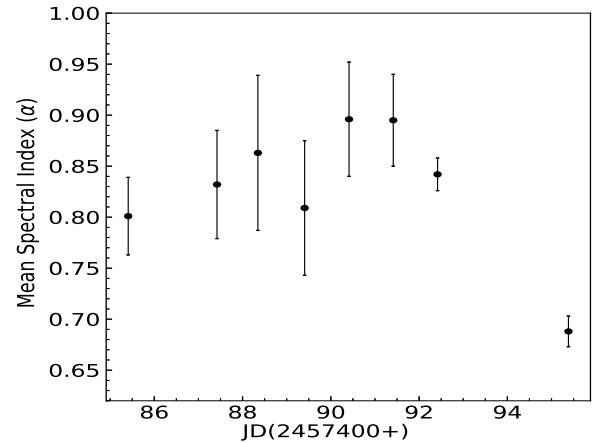}
\caption{Variation of mean optical spectral index with time} 
\label{fig:alpha}
\end{figure}

Since the optical (synchrotron) spectra of blazars are well described by a single power law ($F_{\nu} \propto \nu^{-\alpha}$, where $\alpha$ is the optical spectral index), we extracted the optical spectral energy distributions (SEDs)  across the V, R and I bands to study optical spectral changes in more detail. For this, we have de-reddened the magnitudes in V, R and I bands using the Galactic extinction coefficients ($A_V = 0.142, A_R = 0.113, A_I = 0.078$), taken from the NASA Extragalactic Database (NED\footnote{\url{https://ned.ipac.caltech.edu/}}) and then converted them into $F_{\nu}$. Since the host galaxy contribution for PG 1553$+$113 is negligible \citep[e.g.,][]{1994ApJS...93..125F,2008ApJ...682..775R}, the fluxes were not corrected for a host galaxy.
The optical SED of PG 1553$+$113, in the form of $log(F_{\nu}) - log(\nu)$, is plotted in Fig. \ref{fig:sed} where the flux density ($F_{\nu}$) decreases with the increasing frequency ($\nu$) every night. We fitted the SED with a single power law, in the form of a straight line ($log(F_{\nu}) = -\alpha~log(\nu)+ C$) to get the nightly optical spectral index of the blazar PG 1553$+$113. The results of the fits are given in Table \ref{tab:sed}. 
The mean value of spectral index during our monitoring period is $0.829 \pm 0.017$ and the maximum variation is 0.21, as shown in Fig. \ref{fig:alpha}. These results are close to those found by \cite{1994ApJS...93..125F}.
The temporal variation of V$-$R color indices on STV timescales is also plotted in the bottom panel of Fig. \ref{fig:stv}, which indicates that the color is almost constant during our monitoring period, with a maximum variation of 0.065.

\begin{table}
\caption{Linear fits to color-magnitude plots}            
\label{tab:r_v}                   
\centering    
\resizebox{0.5\textwidth} {!}{  
\hskip-1.cm              
\begin{tabular}{lcccc}           
\hline                		 
Observation date & $m_1^a$ &  $c_1^a$ & $r_1^a$ &  $p_1^a$  \\
yyyy-mm-dd   &         &          &         &                         \\		 
\hline                          
2016-04-06	&$-0.366 \pm  0.253 $ & ~5.521 & -0.273 & 1.600e-01 \\
2016-04-08	&$-0.003 \pm  0.427 $ & ~0.427 & -0.002 & 9.941e-01 \\
2016-04-09	&$-0.674 \pm  0.241 $ & ~9.938 & -0.752 & 3.152e-02 \\
2016-04-10	&$-1.140 \pm  0.556 $ & 16.544 & -0.495 & 6.095e-02 \\
2016-04-11	&$-0.805 \pm  0.611 $ & 11.817 & -0.296 & 2.043e-01 \\
2016-04-12	&$-0.111 \pm  0.425 $ & ~1.959 & -0.060 & 7.974e-01 \\
2016-04-13	&$~0.554 \pm  0.295 $ & -7.529 &  0.426 & 7.828e-02 \\
2016-04-16	&$-0.435 \pm  0.163 $ & ~6.507 & -0.544 & 1.608e-02 \\

\hline                           
\end{tabular}}\\
$^am_1$ = slope and $c_1$ = intercept of CI against R-mag; $r_1$ = Correlation coefficient; $p_1$ = null hypothesis probability
\end{table}


\begin{table}[]
\caption{Straight line fits to optical SEDs}            
\label{tab:sed}                   
\centering     
\resizebox{0.5\textwidth} {!}{ 
\hskip-1.cm                
\begin{tabular}{lcccc}           
\hline                		 
Observation date & $\alpha^a$ &  $C^a$ & $r_2^a$ &  $p_2^a$  \\
yyyy-mm-dd   &         &          &         &                         \\		 
\hline                          
2016-04-06	&$ 0.801 \pm  0.038 $ & -13.033 & -0.999 & 3.026e-02 \\
2016-04-08	&$ 0.832 \pm  0.053 $ & -12.643 & -0.998 & 4.086e-02 \\
2016-04-09	&$ 0.863 \pm  0.076 $ & -12.229 & -0.996 & 5.618e-02 \\
2016-04-10	&$ 0.809 \pm  0.066 $ & -13.079 & -0.997 & 5.165e-02 \\
2016-04-11	&$ 0.896 \pm  0.056 $ & -11.855 & -0.998 & 3.938e-02 \\
2016-04-12	&$ 0.895 \pm  0.045 $ & -11.912 & -0.999 & 3.219e-02 \\
2016-04-13	&$ 0.842 \pm  0.016 $ & -12.743 & -1.000 & 1.183e-02 \\
2016-04-16	&$ 0.688 \pm  0.015 $ & -15.052 & -1.000 & 1.356e-02 \\

\hline                           
\end{tabular}}\\
$^a\alpha$ = spectral index and $C$ = intercept of log($F_{\nu}$) against log($\nu$); $r_2$ = Correlation coefficient; $p_2$ = null hypothesis probability
\end{table}

\section{DISCUSSION AND CONCLUSION} \label{sec:diss}
Studies of flux variability on diverse timescales constitute a powerful method  to better understand the radiation mechanisms of blazars: they provide information about the location, size and dynamics of the emitting regions \citep[e.g.,][]{2003A&A...400..487C}. In blazars, the thermal radiation from the accretion disk is generally overwhelmed by the Doppler-boosted non-thermal radiation from the relativistic jet, so the variability on any measurable timescale is most likely explained by the relativistic jet based models. However, in very low states of blazars, much of the variability in the light curves could be explained by the instabilities in, or hotspots  on, the accretion disks \citep[e.g.,][]{1993ApJ...406..420M,1993ApJ...411..602C}.
Much of longer term blazar variability can reasonably be explained by the shock-in-jet models \citep[e.g.,][]{2015MNRAS.450..541A}. When a shock, assumed to originate from the base of the jet, propagates outward along the relativistic jet, the electrons at the shock front get accelerated to very high energies \citep{1985ApJ...298..114M}. These high energy electrons are then cooled via synchrotron processes while leaving the shock front. Other contributions to blazar variability can arise from wiggles in the jet direction or  helical structures within the jet which cause variations in the Doppler boosting factor \citep[e.g.,][]{1992A&A...255...59C, 1992A&A...259..109G, 1999A&A...347...30V}.   On smaller physical scales and  the shorter timescales observed in this work, relativistic turbulence in the plasma, either crossing a standing shock within the jet which in turn accelerates the electrons to high energies \citep{2014ApJ...780...87M} or otherwise producing fluctuations \citep[e.g.,][]{2015JApA...36..255C,2016ApJ...820...12P}, may dominate.

The spectral or color behavior of blazars can be used to understand the underlying emission mechanisms. Different color behaviours that have been observed in blazars are:  bluer-when-brighter (BWB) and redder-when-brighter (RWB). In some cases, authors have also claimed no clear trend \citep[e.g.,][]{2009ApJ...694..174B,2009ApJS..185..511P}. A BWB trend means the source becomes harder with increasing brightness or softer when its brightness decreases, while a RWB trend indicates opposite behavior. The BWB trend is more commonly observed in  in BL Lac objects, while the FSRQs usually follow a RWB trend \citep[e.g.,][]{2012MNRAS.425.3002G,2015MNRAS.452.4263G}. Nonetheless, it has been found that the same source may follow different trends depending on its variation modes or timescales \citep[e.g.,][]{2003A&A...402..151R,2011MNRAS.418.1640W}.

The commonly observed BWB trend can be interpreted in several different ways that require a great deal of data on many sources to distinguish between them. It may indicate that the two components, one variable ($\alpha_{var}$) and one stable ($\alpha_{const} > \alpha_{var}$), contribute to the overall optical emission with the variable component having a flatter slope than the stable component \citep{2004A&A...419...25F}. The BWB behavior could also be explained using a one component synchrotron model, in which the injection of fresh electrons, having an energy distribution harder than that of the cooled ones, causes an increase in the flux \citep{1998A&A...333..452K,2002PASA...19..138M}. Another possible explanation of BWB chromatism could be the variations in Doppler factor on a ``convex" spectrum caused by the precession of the jet \citep{2004A&A...421..103V}. The RWB chromatism usually observed in FSRQs can arise through the presence of a less variable quasi-thermal emission component from the accretion disk, which can ``contaminate'' the non-thermal jet emission in the rest-frame optical and ultra-violet regions \citep[e.g.][]{2011MNRAS.418.1640W}.

It has been reported in several studies that the amplitude of optical IDV in the LCs of HBLs is statistically significantly smaller than  in the LCs of LBL \citep[][e.g.,]{1998A&A...329..853H,1999A&AS..135..477R,2011MNRAS.416..101G}. The results of our IDV analysis are in line with this conclusion as we saw significant IDV  in only three out of eight R band LCs and in only one out of eight V band LCs. The difference in the optical IDV behaviours of HBLs and LBLs could be due to the stronger magnetic field present in the HBLs \citep{1996ApJ...463..444S} that might prevent the development of features like density inhomogeneities, bends and turbulent eddies in the bases of the jets  \citep{1999A&AS..135..477R}. In particular, it has been argued that an  axial magnetic field $B$ can prevent the formation of instabilities if its value is greater than the critical value $B_c$ given by \citep{1995Ap&SS.234...49R}
\begin{equation}
B_c = \big[4\pi n_e m_e c^2(\gamma^2 - 1)\big]^{1/2} \gamma^{-1},
\end{equation}
where $n_e$ is the local electron density, $m_e$ is the rest mass of electron, and $\gamma$ is the bulk Lorentz factor of the flow.
In HBLs, $B > B_c$ would prevent the development of small-scale structures that can be responsible for the microvariations in the optical light curves when they interact with the relativistic shocks.

We noticed that the variability amplitude decreases with increasing source brightness. On April 8, 12 and 13 the variability amplitudes are 4.76\%, 3.44\%, and 6.17\%, respectively, while the source mean magnitudes are 14.199, 14.181, and 14.246, respectively. This can be explained as the irregularities in a turbulent jet decrease with an increase in source flux, as that increase should arise from a more uniform flow which also reduces the amplitude of variability \citep{2014ApJ...780...87M}.

In the present study of the TeV HBL PG 1553$+$113, made over eight nights in 2016 April with two optical telescopes in India, we found significant IDV flux variations in the R band on three nights  and in the V band only on one of those nights. The photometry was carried out quasi-simultaneously in V and R bands. The blazar did not show large-amplitude variations during our monitoring period.   We detected no strong variations in color with time  nor with brightness during our observations while the mean optical power law spectral index  was $\sim 0.83 \pm 0.02$.  Since our observations were rather short, never exceeding four hours,  it is certainly possible that if we had longer nightly stares at this source we would have seen  more frequent IDV.  Flux variations on an STV timescale were also seen at all three optical wavelengths (R,V, and I), while the colors were found to be almost constant.

An optical photometric study of the BL Lac PG 1553$+$113 was also carried out by \cite{2012MNRAS.425.3002G}. They observed the blazar for IDV on six nights but found no significant IDV or color (B$-$R) variability during any night. On STV timescales they detected genuine flux variability  with no variation in color. \cite{2016MNRAS.458.1127G} also monitored this source for IDV on 7 nights but found significant variations on IDV timescale only on one night. They reported significant flux variability with moderate color variation on STV timescales. We observed the minimum R band magnitude of $R_{mag} = 14.14$ on April 9 which is 0.64 mag fainter than brightest magnitude of  $R_{mag} = 13.5$ mag that was detected during a flaring state by \cite{2006AJ....132..873O}. 
 In addition, we found that the blazar PG 1553$+$113 shows no clear color variation with magnitude,  in accord with earlier measurements of  \cite{2012MNRAS.425.3002G}.

\acknowledgments
We thank the anonymous referee for useful comments and suggestions. We are thankful to Dr.\ Jose Antonio de Diego for a detailed discussion on the nested ANOVA test.\\
\noindent
\software{IRAF (\url{http://iraf.net}), DAOPHOT \RNum{2} \citep{1987PASP...99..191S,1992ASPC...25..297S}, python 2.7 (\url{http://www.python.org})}.

\bibliographystyle{aasjournal}
\bibliography{master}

\end{document}